\documentclass[sigconf]{acmart}

\makeatletter
\def\@ACM@checkaffil{
    \if@ACM@instpresent\else
    \ClassWarningNoLine{\@classname}{No institution present for an affiliation}%
    \fi
    \if@ACM@citypresent\else
    \ClassWarningNoLine{\@classname}{No city present for an affiliation}%
    \fi
    \if@ACM@countrypresent\else
        \ClassWarningNoLine{\@classname}{No country present for an affiliation}%
    \fi
}
\makeatother

\definecolor{mygreen}{RGB}{112, 173, 71}
\definecolor{myred}{RGB}{192, 0, 0}
\definecolor{kellygreen}{rgb}{0.3, 0.73, 0.09}

\newcommand{\cut}[1]{}

\usepackage{amsmath,amsfonts}
\usepackage{amsthm}

\usepackage{algpseudocode}
\usepackage{algorithm}

\usepackage{graphicx}
\usepackage{hyperref}
\usepackage{multirow}
\usepackage{soul}
\usepackage{subcaption}
\usepackage{tabularx}
\usepackage{textcomp}
\usepackage{xcolor}
\usepackage{xspace}
\usepackage{multirow}
\usepackage{booktabs}
\usepackage{graphicx}
\usepackage{threeparttable}  
\usepackage{amsmath}
\usepackage{amsfonts}
\usepackage{tcolorbox}
\usepackage{stmaryrd}
\usepackage{arydshln} 
\usepackage{dashrule}
\usepackage{balance}

\usepackage[normalem]{ulem}
\usepackage{enumitem}

 \newcommand{\sssplit}{{\mathtt{Split}}}
 \newcommand{\ssrecover}{{\mathtt{Recover}}}

\newcommand{\flamingo}{Flamingo\xspace}

\newcommand{\eseafl}{e-SeaFL\xspace}

\newcommand{\node}{\ensuremath{\mathtt{A}}}

\newcommand{\user}{\ensuremath{\mathtt{P}}}

\newcommand{\adv}{\ensuremath{\mathtt{Adv}}}

\newcommand{\khprf}{\ensuremath{\mathtt{KHPRF}}}

\renewcommand{\vec}[1]{\ensuremath{\mathbf{#1}}}

\newcommand{\sss}{\ensuremath{\mathcal{SS}}}

\newcolumntype{C}[1]{>{\centering\arraybackslash}p{#1}}

\newcommand{\cnaf}{\ensuremath{\mathsf{AF_{FL}}}\xspace}

\AtBeginDocument{%
  \providecommand\BibTeX{{%
    \normalfont B\kern-0.5em{\scshape i\kern-0.25em b}\kern-0.8em\TeX}}}

\copyrightyear{2025}
\acmYear{2025}
\setcopyright{cc}
\setcctype{by}
\acmConference[WiSec 2025]{18th ACM Conference on Security and Privacy in Wireless and Mobile Networks}{June 30-July 3, 2025}{Arlington, VA, USA} 
\acmBooktitle{18th ACM Conference on Security and Privacy in Wireless and Mobile Networks (WiSec 2025), June 30-July 3, 2025, Arlington, VA, USA} 
\acmDOI{10.1145/3734477.3734719} 
\acmISBN{979-8-4007-1530-3/2025/06}

\begin{document}

\title{Standing Firm in 5G: A Single-Round, Dropout-Resilient Secure Aggregation for Federated Learning}


\author{Yiwei Zhang}
\affiliation{%
  \institution{Purdue University}
}
\email{yiweizhang@purdue.edu}

\author{Rouzbeh Behnia}
\affiliation{%
  \institution{University of South Florida}
}
\email{behnia@usf.edu}

\author{Imtiaz Karim}
\affiliation{%
  \institution{Purdue University}
}
\email{karim7@purdue.edu}

\author{Attila A. Yavuz}
\affiliation{%
  \institution{University of South Florida}
}
\email{attilaayavuz@usf.edu}

\author{Elisa Bertino}
\affiliation{%
  \institution{Purdue University}
}
\email{bertino@purdue.edu}

\renewcommand{\shortauthors}{Trovato and Tobin, et al.}

\begin{abstract}
Federated learning (FL) is well-suited to 5G networks, where many mobile devices generate sensitive edge data. Secure aggregation protocols enhance privacy in FL by ensuring that individual user updates reveal no information about the underlying client data. However, the dynamic and large-scale nature of 5G-marked by high mobility and frequent dropouts-poses significant challenges to the effective adoption of these protocols.  Existing protocols often require multi-round communication or rely on fixed infrastructure, limiting their practicality.
We propose a lightweight, single-round secure aggregation protocol designed for 5G environments. By leveraging base stations for assisted computation and incorporating precomputation, key-homomorphic pseudorandom functions, and $t$-out-of-$k$ secret sharing, our protocol ensures efficiency, robustness, and privacy. Experiments show strong security guarantees and significant gains in communication and computation efficiency, making the approach well-suited for real-world 5G FL deployments.
\end{abstract}

\keywords{5G; Federated Learning; resilient networks; privacy}

\ccsdesc[500]{Security and privacy~Security protocols}
\ccsdesc[300]{Security and privacy~Privacy protections}

\newcommand{\imtiaz}[1]{\textbf{\color{purple}Imtiaz: #1}}





\maketitle

\section{Introduction}

Federated learning (FL) has emerged as a powerful approach for enabling privacy-preserving machine learning across distributed devices, allowing users to collaboratively train models without sharing raw data~\cite{mcmahan2017communication}. While FL has been deployed over conventional communication infrastructures, the rapid rollout of fifth-generation (5G) networks introduces both opportunities and new challenges. 
On the one hand, 5G offers high bandwidth, low latency, and support for massive device connectivity—features that make it well-suited for large-scale, on-device FL deployments. Real-world applications already highlight FL’s potential in 5G settings, such as real-time analytics for connected vehicles and privacy-preserving medical diagnostics~\cite{perifanis2023federated,liu2020secure}. FL can also benefit network providers by improving tasks like channel estimation~\cite{li2019deep}, thereby enhancing overall network performance while maintaining user privacy.
On the other hand, these same 5G features create a highly dynamic environment~\cite{jabbar2011survivable}. Devices frequently join and leave the network, connectivity conditions fluctuate, and the scale of participation can surpass what existing FL protocols were designed to handle.

These 5G-specific conditions pose significant obstacles for secure aggregation~\cite{bonawitz2017practical}, a critical component of FL that ensures individual model updates remain private during server-side aggregation. Most existing secure aggregation techniques assume relatively stable device participation and moderate network scale, often relying on multiple rounds of interaction or fixed infrastructure. In contrast, a practical secure aggregation protocol for 5G must operate efficiently under high churn, large user populations, and minimal communication rounds-all while preserving strong security guarantees.

\noindent
\textbf{Requirements of 5G for FL.}
5G’s capacity to connect large numbers of devices simultaneously greatly 
expands the pool of potential FL participants, but it also introduces significant communication and computation overhead if aggregation involves multiple rounds of interaction. 
Meanwhile,
5G networks exhibit high mobility and signaling, leading to unpredictable connectivity and higher churn rates, which can degrade the reliability and speed of FL aggregation. 
Therefore, 
secure aggregation solutions in 5G must satisfy: 

\begin{itemize}[leftmargin=*] 
\item \textbf{Scalability.} 
Secure aggregation introduces additional computation and communication overhead in addition to the local training on mobile devices. Therefore, given the resource constraints of these battery-powered devices, this can directly impact the performance and scalability of these systems. 
Since each user’s training parameters must be collected, the overhead of computation and communication should remain manageable even when thousands or millions of devices are involved. 
\item \textbf{Resilience.} Given the dynamic nature of 5G, where user devices (e.g., cell phones) and even the base stations can go offline unexpectedly, the secure aggregation mechanism must be robust against dropouts. 
\item \textbf{Compatibility.} To enable seamless deployment, the protocol should be compatible with existing 5G infrastructures without demanding substantial hardware or protocol modifications. 
\end{itemize}

\noindent
\textbf{Current Secure Aggregation Solutions.}
Existing approaches to secure aggregation primarily rely on secure multi-party computation (MPC)~\cite{ma2023flamingo,guo2024microsecagg} or pairwise masking~\cite{behnia2023efficient,karthikeyan2024opa,bell2020secure}. For instance, \flamingo~\cite{ma2023flamingo} employs a multi-round protocol where each user reuses a secret to generate masks; a subset of users (i.e., decryptors) then interact over multiple rounds to reveal active user masks and eliminate offline user contributions. 
\eseafl~\cite{behnia2023efficient}, based on pairwise masking, reduces the communication overhead by introducing assisting nodes that hold shared secrets, achieving single-round secure aggregation. 
OPA~\cite{karthikeyan2024opa} also follows the pairwise masking paradigm, but instead of relying on pre-shared secrets, it uses key-homomorphic cryptography. Users interact with a lightweight committee to remove global masks by exchanging fresh auxiliary information in each training round.

However, these methods face limitations under 5G conditions. While \flamingo can handle dropouts via secret sharing, its multi-round interactions significantly increase communication overhead. \eseafl relies on a designated set of assisting nodes, and if these nodes go offline unexpectedly, it can cause the entire protocol to fail. Both protocols have stringent requirements and rely on complex public key infrastructures (PKIs), making their deployment infeasible in fast-evolving 5G networks.
OPA adopts a stateless, one-shot design without pre-shared secrets, but supports only one training round per execution. To run multiple iterations-as is common in FL-it must restart with new auxiliary exchanges each round, incurring significant overhead.
Moreover, OPA lacks precomputation mechanisms to alleviate the computational and communication overhead of its single-round approach, making it less suitable for resource-constrained 5G environments.

\vspace{5pt}
\noindent
\textbf{Our Contributions.}
To address these challenges, we propose a novel secure aggregation framework tailored to FL in large-scale 5G environments. 
Our approach leverages base stations to assist the server in aggregating user updates to significantly reduce the overall communication and computation overhead, achieving a \emph{single-round} secure aggregation protocol. We also introduce a simple but effective pre-computation method to meet the stringent performance requirements of mobile devices in 5G networks. By incorporating key-homomorphic pseudo-random functions (KHPRFs) and a robust $t$-out-of-$k$ secret sharing scheme, our protocol tolerates both user equipment and base station dropouts without compromising security or correctness. Crucially, our solution integrates seamlessly with standard 5G architectures and does not impose significant additional infrastructure requirements. In summary, our main contributions include: 

\begin{itemize}[leftmargin=*] 
\item A \emph{single-round} secure aggregation protocol optimized for large-scale 5G settings, supported by \emph{precomputation} techniques and base stations to minimize computation and communication overhead and the risk of data leakage.
\item Resilience mechanisms that ensure continuous operation despite unexpected device or base station dropout, thus maintaining reliable model convergence in 5G deployments.
\item A thorough security analysis and experimental evaluation showing the computational and communication efficiency of our approach, as well as its robust security guarantees. 
\end{itemize}

\section{Preliminaries and Models}

Following~\cite{bell2020secure, ma2023flamingo}, we adopt the definitions of privacy for secure aggregation protocols as follows. 

\begin{definition}[Key-Homomorphic Pseudorandom Function (KHPRF)~\cite{boneh2013key}]\label{def:khprf}
A key-homomorphic pseudorandom function is \(\khprf: \mathcal{K} \times \mathcal{M} \rightarrow \mathcal{Y}\),
where \(\mathcal{K}\) is the key space, \(\mathcal{M}\) is the domain of inputs, and \(\mathcal{Y}\) is the output space. This function satisfies:
1) \textit{Key-Homomorphism}: For any keys \(k_1, k_2 \in \mathcal{K}\) and any input \(m \in \mathcal{M}\), \(\khprf(k_1 + k_2, m) = \khprf(k_1, m) + \khprf(k_2, m)\)
    where the addition operations are performed in the appropriate key and output groups. 
    This property naturally extends to any finite sums of keys, i.e., \(\khprf(\sum_{i\in I}k_i, m) = \sum_{i\in I}\khprf(k_i, m)\),
    for any index set \(I\subseteq \{1,\dots,n\}\).
2) \textit{Pseudorandomness}: For a uniformly chosen \(k \in \mathcal{K}\), the function \(\khprf(k,\cdot)\) is computationally indistinguishable from a truly random function mapping \(\mathcal{M}\) to \(\mathcal{Y}\).

\end{definition}

\begin{definition}[\(t\)-out-of-\(n\) Secret Sharing \cite{shamir1979share}]\label{def:ss}
A secret sharing scheme \(\sss = \{\sss.\sssplit, \sss.\ssrecover\}\) for threshold \(t\) out of \(n\) operates over a message space \(\mathcal{M}\). It consists of two algorithms:
\begin{itemize}
    \item \(\{s_1,\dots,s_n\} \gets \sss.\sssplit(s)\): 
    Given a secret \(s \in \mathcal{M}\), the algorithm outputs \(n\) shares \(\{s_1,\dots,s_n\}\). Any subset of at least \(t\) shares is sufficient to recover \(s\), whereas any subset of fewer than \(t\) shares reveals no information about \(s\).
    \item \(s \gets \sss.\ssrecover(\mathcal{S'})\): 
    Given any subset \(\mathcal{S'} \subseteq \{s_1,\dots,s_n\}\) of size \(\lvert \mathcal{S'} \rvert \ge t\), the algorithm reconstructs and outputs the secret \(s\). If \(\lvert \mathcal{S'} \rvert < t\), it outputs \(\bot\) (failure).
\end{itemize}
\end{definition}

\begin{definition}[$\alpha$-summation ideal functionality \cite{bell2020secure}]\label{def:alphasummation}
Given $p,n,d$ as integers and consider a set $L\subseteq [n]$ with associated data vectors $\mathcal{W}_L:=\{\Vec{w}_i\}_{i\in {L}}$ where $\Vec{w}_i \in\mathbb{Z}_p^d$. 
Given a threshold $0 \leq \alpha \leq 1$ and $Q_{L}$ as the set of partitions of $L$  and a set of pairwise disjoint subsets  $ \{L_{1}, \dots, L_l\} \in Q_{L}$, the $\alpha$-summation ideal functionality $\mathcal{F}_{\Vec{w},\alpha}(\cdot)$
computes $\mathcal{F}_{\vec{w},\alpha}(\{L_i\}_{i\in[1,\dots,l]}) \rightarrow  \{\Vec{s}_i\}_{i\in[1,\dots,l]}$,
where 
\begin{equation*}
    \forall j\in[1,\dots,l] ,\Vec{s}_{j} =
    \begin{cases}
       \sum_{j\in Q_L}\vec{w}_j & \text{if}~Q_L|\geq \alpha|L| \\
      
      \bot & \text{else.}
    \end{cases}    
\end{equation*}

\end{definition}

\begin{definition}[Privacy of Secure Aggregation \cite{bell2020secure}]\label{def:aggSec}

Let $\Sigma$ be a Shamir secret sharing protocol instantiated with a security parameter $\kappa$, and $\mathcal{F}$ be an ideal functionality that only outputs the aggregation if at least $\alpha n$ defined as honest users are participating, where $0 <\alpha \leq 1$. 
An aggregation protocol $\Gamma$ is said to preserve privacy against an adversary $\mathcal{A}$ if there exists a probabilistic polynomial-time (PPT) simulator $\mathtt{Sim}$ such that, for any iteration $t \in [T]$ and any set of input vectors $\mathcal{W}^t = \{\Vec{w}_1, \ldots, \Vec{w}_n\}$, the output produced by $\mathtt{Sim}$ is computationally indistinguishable from the adversary's view. 
The adversary $\mathcal{A}$ is assumed to have control over $NC$, a $\lambda_\user$ fraction of users $U_c$, and a $\lambda_\node$ fraction of assisting nodes $BS_c$. Its view comprises the combined view of the compromised server $NC^*$, the set of compromised users, and the set of compromised assisting nodes:
$
\mathsf{Real}^{\Gamma}(\adv, \{\Vec{w}_i\}_{i\notin U_c})
\approx_\kappa
\mathsf{Simul}^{\Gamma, \mathcal{F}_{\mathcal{W}^t,\alpha}(\cdot)_{\{\Vec{w}_i\notin U_c\}{}}}(\adv) 
$.
\end{definition}

\subsection{System Model}

\begin{figure}[t]
  \centering
  \includegraphics[width=0.76\linewidth]{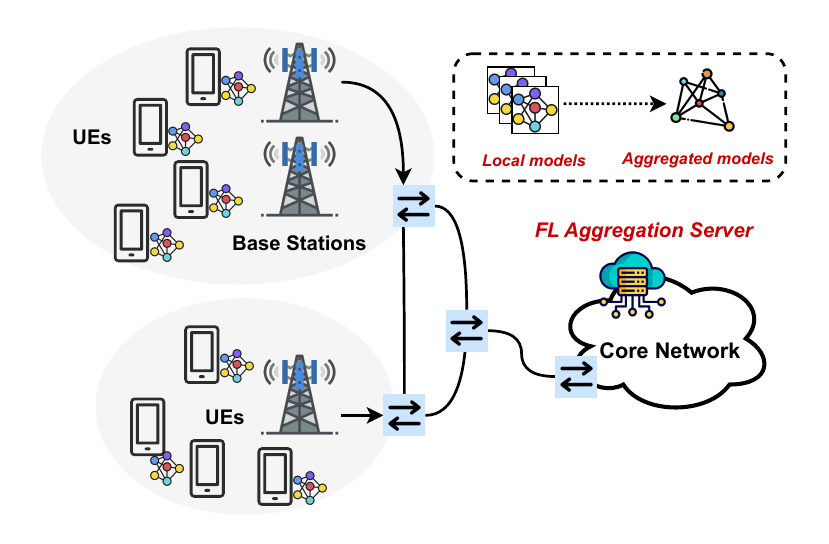}
  \caption{5G+FL Framework.}
  \label{fig:framework}
\end{figure}

\noindent \textbf{5G-Integrated FL Ecosystem.}
As illustrated in Figure~\ref{fig:framework}, our architecture comprises three core components: \emph{user equipment (UE)}, \emph{base stations (BSs)}, and the \emph{core network (CN)}.
It harnesses 5G’s high bandwidth, low latency, edge-processing capabilities as well as security features to support efficient, privacy-preserving model training across diverse user devices.

UEs (e.g., mobile or IoT devices) collect local data and train models at the network edge. They typically possess basic edge-computing capabilities, allowing them to perform local learning without uploading raw data to the aggregation server and masking their local updates before transmission.
Each UE operates under a BS for connectivity; BSs often possess edge-computing resources to help with partial unmasking and handling potential device dropouts. 
The CN orchestrates system-wide coordination and security. Typically, the \emph{Authentication Server Function} (AUSF) in the CN ensures secure device authentication, while the \emph{Application Function} (AF), either hosted in the CN or provided by an external third party, delivers user-facing services.
When integrating with FL functionality, a dedicated AF module (\cnaf) coordinates the end-to-end FL process by collecting UE updates, aggregating them into a global model, and distributing the result to participating UEs. 
Robust end-to-end encryption and integrity checks among the three components are enforced via the 5G-AKA protocol~\cite{3gppakmastudy}.

\noindent \textbf{FL Workflow.} 
The FL procedure comprises two main phases. 
In the one-time \emph{Setup Phase}, cryptographic materials are configured for UEs and BSs, including establishing secret shares to protect local updates and tolerate BS dropouts. 
In the iterative \emph{Aggregation Phase}, UEs independently train local models, mask their updates, and send them to \cnaf. The \cnaf then unmasks and aggregates these updates to compute a global model, which is subsequently distributed back to the UEs for further local training.



\subsection{Threat Model} 
Based on the 3GPP Technical Specifications~\cite{3gppakma,3gppakmastudy,3gppsys} and relevant literature~\cite{liu2020secure,tran2019federated}, we consider the following adversarial behaviors and practical constraints within our 5G-based FL framework:

\begin{itemize}[leftmargin=*] 
\item 
We assume that the CN in 5G is generally trusted, with the exception of the \cnaf responsible for model aggregation in FL. Since \cnaf may be provided by a third-party service, it can behave maliciously, attempting to infer private user data or conducting membership inference attacks based on aggregated updates~\cite{carlini2021extracting, shokri2017membership, carlini2019secret}. In contrast, other CN components (e.g., AUSF) are considered benign, as they typically reside within the trusted 5G Core Network infrastructure \footnote{We note that we can lift these trust assumptions on CN and AUSF by assuming a simple digital signature scheme akin to the malicious model in \cite{behnia2023efficient}.}.



\item 
UEs may drop out during FL training due to network instability, node mobility, or active adversarial efforts~\cite{lichtman20185g, nyangaresi2022optimized}. Their unpredictable mobility across different BSs can further complicate model consistency. 
Likewise, BSs are susceptible to outages or unavailability caused by network failures, targeted attacks, or other system constraints~\cite{jabbar2011survivable}. 
\item 
To leverage the existing 5G security mechanisms~\cite{3gppakma} rather than introducing new ones (such as PKI), we consider that the FL process begins only after the 5G-AKA procedure is completed. This ensures that communication keys (e.g., long-term keys and session keys) have been securely negotiated and distributed, thus encrypting and protecting the integrity of all transmitted information among UEs, BSs, and CN (AUSF and \cnaf). Consequently, even if adversaries intercept communication channels, they cannot decrypt or forge valid messages unless they obtain direct access to a compromised user device\footnote{Similar to previous work on secure aggregation \cite{ma2023flamingo,behnia2023efficient} we do not consider impersonated or compromised UEs and BSs as our paper focuses on secure aggregation instead of data poisoning.}.


\end{itemize}

\section{Our Protocol}
\begin{figure}[t]
  \centering
  \includegraphics[width=0.88\linewidth]{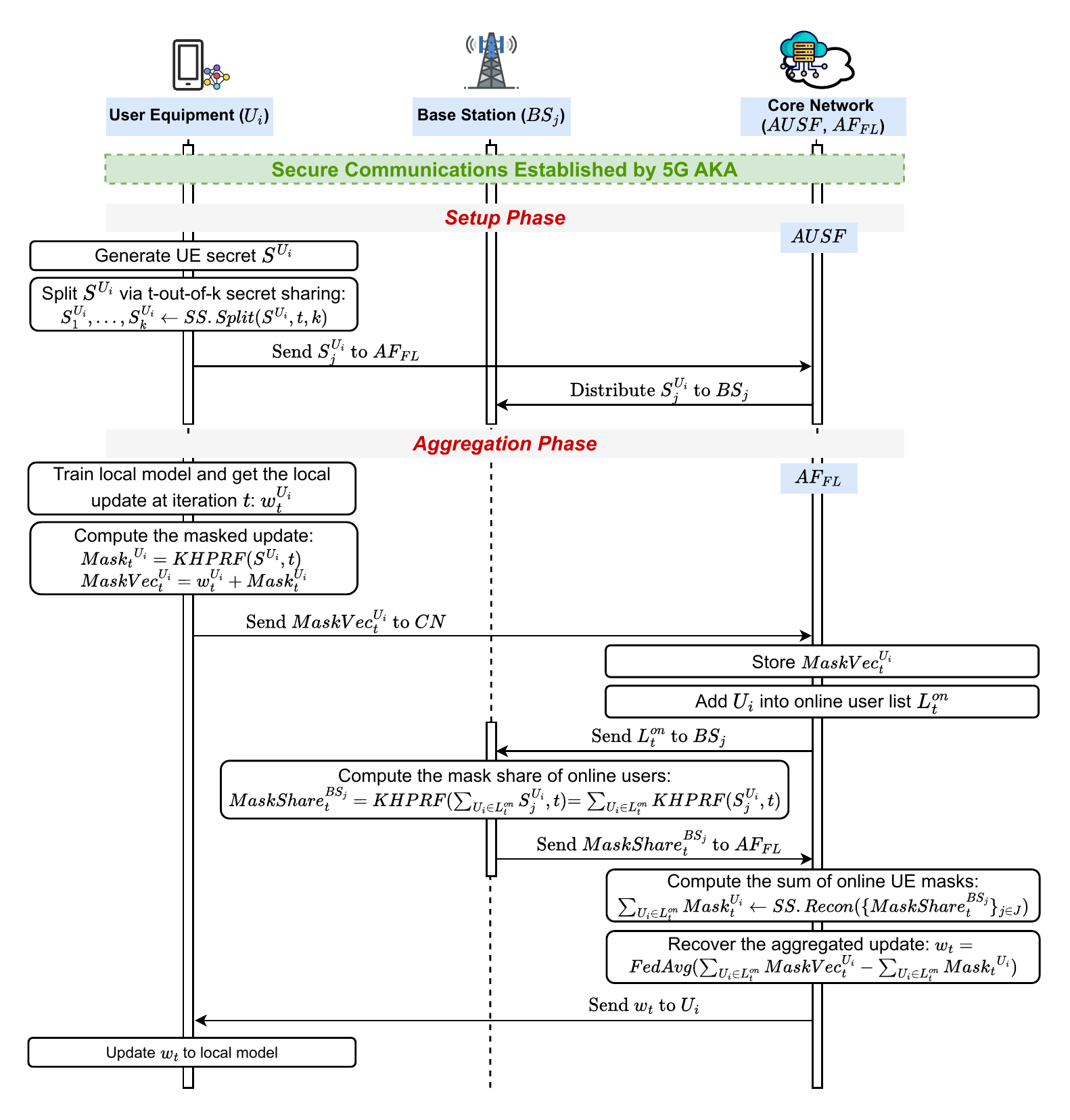}
  \caption{5G-based FL secure aggregation protocol.} \label{fig:protocol}
\end{figure}

We now describe our 5G-based FL secure aggregation protocol (see Figure~\ref{fig:protocol}).
Our protocol is designed to meet the high connectivity, mobility, resilience and compatibility demands of 5G. By combining threshold secret sharing, KHPRFs, and BS-assisted unmasking, it protects individual user data from inference attacks and remains robust against UE and BS dropouts. Each UE and BS has single-round communication with the CN, maintaining high efficiency. 

Our system consists of three primary entities: 
1) $n$ UEs (denoted by ${U_i}$ where $i\in\{1, ..., n\}$); 
2) $k$ BSs (denoted by ${BS_j}$ where $j\in\{1, ..., k\}$) in a certain region;
and 3) a single CN that hosts an AF as the FL aggregation server (\cnaf) as well as an AUSF to forward sensitive messages.
The protocol operates in two main phases: an initial \emph{Setup} phase (one-time) to establish cryptographic secrets, followed by a single-round \emph{Aggregation} phase (iterative) for secure model updates.


\subsection{Setup Phase}
The Setup phase is executed once to bootstrap trust and secrets among the UEs and the BSs, ensuring resilience against BS dropouts and enabling secure masked aggregation.

\noindent \textbf{Step 1.1: UE secret generation and distribution.}
Each UE initializes a secret $S^{U_i}$. 
To facilitate the recovery of user masks during aggregation,
this secret $S^{U_i}$ is partitioned into $k$ splits, $\{S^{U_i}_j, \forall j \in [k]\}$, 
using a $t$-out-of-$k$ secret sharing scheme $\sss.\sssplit(\cdot)$.
The UE sends these shares to the CN via its currently assigned BS.
The AUSF within CN then forwards and distributes one share to each BS in the designated region.
This ensures that the UE secret can be reconstructed if at least $t$ shares are collected, thereby mitigating the impact of BS dropouts.

\subsection{Aggregation Phase}
At each iteration $t$, the protocol executes three steps to securely aggregate local updates and refresh the global model.

\noindent \textbf{Step 2.1: Local model training and masked update generation.} 
Each $U_i$ trains a local model on its private dataset and obtains a parameter update\footnote{The protocol accommodates both gradient- and parameter-based FL variants.} $\Vec{w}_t^{U_i}$.
To protect $\Vec{w}_t^{U_i}$ from direct exposure to the \cnaf, $U_i$ computes a mask vector using the KHPRF: \(\small Mask_{t}^{U_i} = KHPRF(S^{U_i}, t)\).
The final masked update is \(\small MaskVec_{t}^{U_i} = \Vec{w}_t^{U_i} + Mask_{t}^{U_i}\), which is then sent to to the \cnaf.

Since the mask is derived solely from $S^{U_i}$ and the known iteration index $t$, our protocol adopts a \textbf{pre-computation} strategy. That is, a UE can precompute all necessary masks (such as $T$ masks for the first $T$ iterations) before the runtime aggregation phase, minimizing on-the-fly computation costs.

\noindent \textbf{Step 2.2: Online user registration and mask share computation.} 
Once the \cnaf receives $MaskVec_{t}^{U_i}$, it logs $U_i$ in the \textit{online UE list} $L_t^{on}$, identifying UEs that remain active for iteration $t$. 
The \cnaf checks that $|L_t^{on}| \ge \alpha n$; if so, it shares this list with all BSs (in a certain region).

Each $BS_j$ computes its \textit{mask share} by summing the secret shares received from all UEs in $L_t^{on}$ and applying the KHPRF:
\(\small MaskShare_{t}^{BS_j}\)
\(= \khprf(\sum_{U_i \in L_t^{on}} S_j^{U_i}, t) \
= \sum_{U_i \in L_t^{on}} \khprf(S_j^{U_i}, t)\).
Owing to the key-homomorphic property, the final output is equivalent to the sum of individual $\khprf(\cdot)$ evaluations, enabling the \cnaf to only recover the \emph{aggregated mask} while never seeing any single user’s update and mask.

\noindent \textbf{Step 2.3: Secure aggregation and global model update.} 
Using the masked UE updates and the mask shares,
the \cnaf first reconstructs the aggregated mask of online UE masks via the $t$-out-of-$k$ secret reconstruction algorithm:
\(\sum_{U_i \in L_t^{on}} Mask_{t}^{U_i} = \sss.\ssrecover(\)
\(\{{MaskShare}_{t}^{BS_j}\}_{j\in J})\), where $J$ is a subset of BSs which are online.


The \cnaf then computes the global aggregated update $\Vec{w}_t$ 
by subtracting the sum of masks from the sum of all masked UE updates, then applying a standard FL aggregator (e.g., FedAvg~\cite{mcmahan2017communication}): \(\small \Vec{w}_t = \mathtt{FedAvg}(\sum_{U_i \in L_t^{on}} MaskVec_{t}^{U_i} - \sum_{U_i \in L_t^{on}} Mask_{t}^{U_i})\).
This sum of masks can make sure to “unmask” the global update without revealing individual user parameters, i.e., only the aggregated result $\Vec{w}_t$ is revealed, preventing the \cnaf from accessing any user-level update directly. 

After computing the global update $\Vec{w}_t$, the \cnaf then propagates it to all participating UEs. 
UEs then incorporate these aggregated parameters into their local models, completing the $t$-th iteration. 
Notably, new or reconnected UEs can join seamlessly after completing their own setup phase, while the aggregation phase continues iteratively until the global model converges or meets predefined performance criteria.

\subsection{Security Analysis}
\begin{theorem}
    The protocol proposed above with $n$ users, $k$ base stations and the aggregation server \cnaf is private against an adversary $\mathcal{A}$ where $\mathcal{A}$ is able to compromise $(1-\alpha)n $ users, $k-1$ base stations and the aggregation server \cnaf. 
\end{theorem}

\begin{proof}
    Following Definition~\ref{def:aggSec}, we consider a simulator $\mathcal{C}$ and establish indistinguishability through a standard hybrid argument. This is done by presenting a set of successive hybrids that are computationally indistinguishable. We start by defining the behavior of $\mathcal{C}$ during the Setup and Aggregation phases. 
    In the Setup phase, honest users $U_i$ compute their secret shares $\{S^{U_i}_1 \dots, S^{U_i}_k\}$ and encrypt them (using AKA-ENC) for the target $BS$. The encrypted shares are distributed to the base stations via AUSF. 
    During the aggregation phase, users compute their masked update $MaskVec_t^{U_i}$ and send it to \cnaf. Next, \cnaf adds the user to an authenticated list of participating users. We now present our hybrids. 
    
    \noindent
    $\mathtt{Hyb0}$: This hybrid represents the actual execution of the protocol, during which  $\mathcal{A}$ interacts with the honest entities.  
    
    \noindent
    $\mathtt{Hyb1}$: $\mathcal{C}$ is introduced. $\mathcal{C}$ is assumed to possess all the secrets of the honest parties.
    
    \noindent
    $\mathtt{Hyb2}$: The behavior of the honest parties $U_i $ and base stations $BS_j$ is modified by selecting a random shared secret key from the key space $\mathcal{K}_\mathtt{}$ instead of executing   $\sss.\sssplit(\cdot)$ algorithm. The security of the secret sharing protocol (e.g., as in~\cite{shamir1979share,bell2020secure}) and the encryption scheme ensure the indistinguishability of this hybrid from the previous one. 
    
    \noindent
    $\mathtt{Hyb3}$: Each honest user $U_i$ replaces the masked update $MaskVec_t^{U_i}$ sent to \cnaf with a random vector $MaskVec_t’^{U_i}$. Given our protocol requires at least one base station to be honest, the indistinguishability of this hybrid is ensured, as $\mathcal{A}$ does not have knowledge of at least one honest base station. Therefore, in $\mathsf{Real}$, the masked update follows the same distribution as $MaskVec_t’^{U_i}$ in $\mathsf{Simul}$. 
    
    \noindent
    $\mathtt{Hyb4}$: The aggregated masking term  $MaskShare_t^{BS_j}$ outputted by the honest base stations $BS_j$ is replaced with $MaskShare_t’^{BS_j}$  by using the ideal functionality $\mathcal{F}_{\Vec{w}, \alpha}(L_t'^{on})$ where $L_t'^{on}$ is the list of honest participating users. Since $\mathcal{A}$  has no knowledge of the honest entities’ shared secret,  the distribution of the aggregated masking term in $\mathsf{Simul}$ is identical to that in $\mathsf{Real}$ and the view of this hybrid remains indistinguishable.  
    
    \noindent
    $\mathtt{Hyb5}$: $\mathcal{C}$ sends the output of the ideal functionality as the intermediate model  $\Vec{w}_t$. Note that the ideal functionality does not return $\bot$, given the condition on the fraction of honest users. Consequently, this hybrid remains indistinguishable from the previous one, as the local updates are unknown to $\mathcal{A}$ in $\mathsf{Real}$.

 From the above, we have demonstrated that the view of all corrupted parties under the control of $\mathcal{A}$ is computationally indistinguishable from their view in $\mathsf{Real}$.
\end{proof}
\section{Evaluation}
We evaluate our 5G-based FL secure aggregation protocol by comparing it against state-of-the-art alternatives. The goal is to demonstrate the efficiency, communication overhead, and resilience of our approach in realistic 5G network scenarios.


\subsection{Experiment Setup}
\noindent \textbf{Implementation}
We implemented our protocol in Python (approximately 1,200 lines of code) using the ABIDES~\cite{byrd2019abides} 
simulation framework,
which
enables controlled testing of multi-round FL aggregation protocols, allowing us to replicate realistic 5G network behaviors in a simulated environment. We used ASCON~\cite{dobraunig2021ascon} as the pseudo-random function (PRF) for mask generation, given its lightweight design and suitability for resource-constrained devices.

\noindent \textbf{Experiment Environments.}
All experiments were conducted on an x86\_64 Linux server equipped with an AMD Ryzen Threadripper PRO 5965WX (24-core CPU), 256 GB of RAM, and three NVIDIA GeForce RTX 3090 GPUs for model training. We do not include model-training overhead in our measurements, as our primary focus is the communication and computation costs introduced by the secure aggregation protocol itself.

\noindent\textbf{Evaluation Metrics.} We assess our protocol using two primary metrics.
\textit{Efficiency} is quantified by computation time and communication overhead during both the setup and aggregation phases. 
\textit{Resilience} is measured by the ability of the protocol to maintain global model accuracy in the presence of UE and BS dropouts.

\noindent\textbf{Baselines.} 
We compare our protocol with two secure aggregation protocols, \eseafl and \flamingo, as they have a certain ability to be resistant to device dropout.
Under similar capabilities, we record the computation time and communication overhead to evaluate the efficiency of our protocol.
For resilience evaluation, we include a baseline variant of our protocol that simply halts FL updates (i.e., distribute the original global model update) if the aggregated update cannot be correctly reconstructed.

\noindent\textbf{Parameter Selection.} 
For a fair comparison, all protocols are evaluated under the same conditions. 
Unless otherwise stated, our experiments use four BSs ($k=4$), and eight UEs ($n=8$). The dropout rate is set to $\alpha = \tfrac{1}{3}$, meaning up to one-third of the UEs and BSs may disconnect during training. 
A $(t,k)$ threshold with $t=(1-\alpha)\cdot k$ (i.e., $t=3$), ensuring successful reconstruction of the aggregated mask as long as at least three BSs remain online.
Moreover, we consider simulated dropout scenarios by reducing the number of UEs from 8 to 2 and BSs from 4 to 2, running tests over 10 training iterations. The key evaluation metric is the accuracy of the global model, which measures the impact of dropouts on FL performance, allowing us to analyze the robustness of the protocol under varying dropout conditions.

\subsection{Results}

\subsubsection{Efficiency}
\begin{figure}[t]
  \centering
  \includegraphics[width=0.84\linewidth]{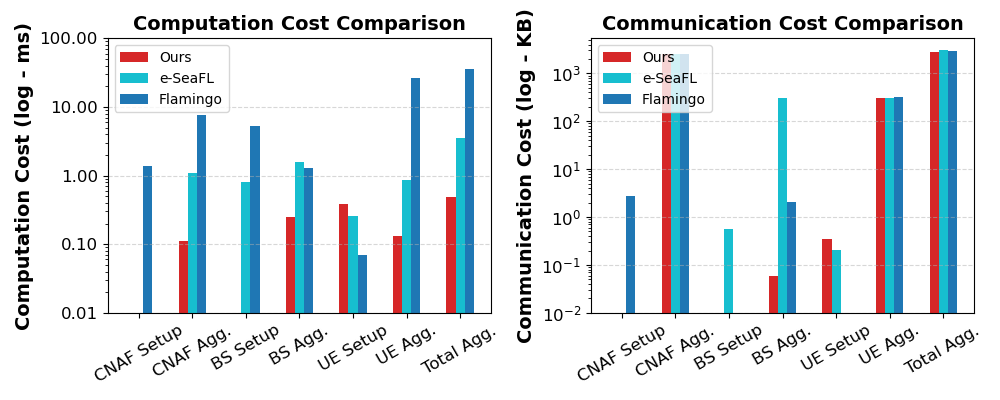}
  \caption{Efficiency Comparison.}
  \label{fig:cost}
\end{figure}

Figure~\ref{fig:cost} reports the computation and communication overhead of our solution compared with \eseafl and \flamingo. Overall, our protocol demonstrates significantly lower aggregation‐phase latency and competitive communication costs.

\noindent \textbf{Setup phase.}
During the one-time setup, our protocol imposes minimal load on both the \cnaf and the BSs, as most computational tasks are offloaded to the UEs (e.g., precomputation). Although this leads to a slightly higher setup cost on the UE side compared to \eseafl and \flamingo, the precomputation of masks at this phase substantially reduces overheads during aggregation. 
In contrast, \flamingo and \eseafl have to perform distributed key generation or shared secret computations among the BSs and the \cnaf, which increase the setup cost on those devices.

\noindent \textbf{Aggregation phase.}
Once FL procedure has been initialized, our protocol achieves markedly faster secure aggregation than \eseafl and \flamingo. 
This improvement is primarily due to: (i) shifting the bulk of cryptographic tasks to a one-time precomputation step in the setup phase; and (ii) employing a key‐homomorphic framework that avoids per‐BS decryption operations when handling dropped or offline clients. 
In comparison, \flamingo tends to incur significant additional steps to manage these dropouts at the BS level (such as decryption), while \eseafl, without a precomputation step, requires each BS to compute large mask vectors at each iteration. 

In terms of communication cost, our protocol remains on par with or improves with respect to \eseafl and \flamingo. 
On the UE and the \cnaf side, our protocol has a similar communication throughput of \eseafl, due to the masked UE updates and aggregated final update, as well as an online user list in our protocol.
Notably, BSs in our scheme benefit from transmitting far fewer bits, since each device only sends an aggregated mask share rather than a full vector.
The mask shares from the BSs can be expanded on the \cnaf.
\flamingo incurs more communication cost due to its multiple interactions between \cnaf and BSs for dropout management, like online UE/BS lists and encryption/decryption results, which create additional communication steps that our approach avoids. 

\subsubsection{Resilience to UE and BS Dropout}

\begin{figure}[t]
  \centering
  \includegraphics[width=0.84\linewidth]{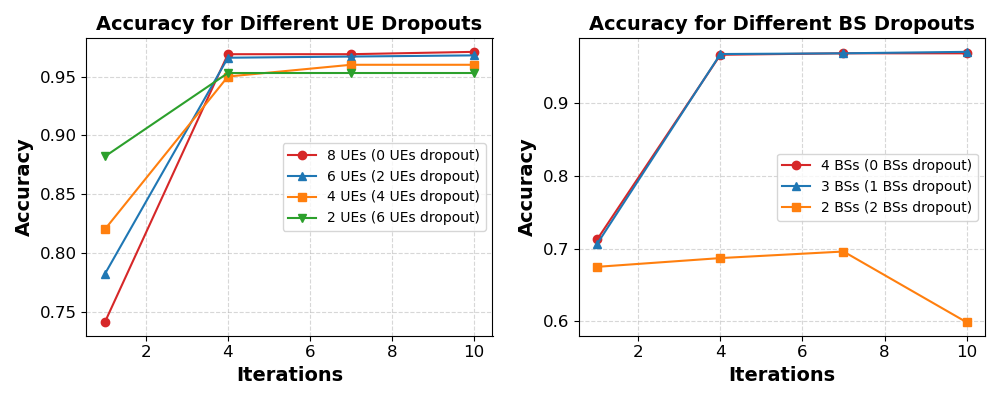}
  \caption{Dropout Resilience.}
  \label{fig:dropout}
\end{figure}

Given the distributed nature of FL, real-world deployments often encounter scenarios where participating nodes may become temporarily unavailable due to network failures, device disconnections, or power constraints.
Figure~\ref{fig:dropout} illustrates the impact of UE and BS dropouts on the global model accuracy.

The left panel of Figure~\ref{fig:dropout} shows the impact of varying UE dropouts on the global model accuracy. We can observe that even as the number of dropped UEs increases, the overall accuracy remains relatively stable. 
This resilience is due to the effective reconstruction of aggregated updates via active BSs.
However, when a larger fraction of UEs drop out, the final accuracy experiences a slight decline due to the reduction in the amount of data contributing to the model update, which also violates the privacy guarantee~\ref{def:alphasummation}.

The right panel shows the impact of BS dropouts on accuracy. When up to one BS drops out, the model maintains a high accuracy level. However, when two BSs drop out, the accuracy decreases significantly. 
This is expected, as our $(3,4)$ secret sharing scheme in the experiment requires at least three BSs to reconstruct the aggregated mask correctly.
With only two BSs remaining, the server is unable to aggregate UE contributions effectively, preventing meaningful global model updates and leading to a stagnation in accuracy at a lower level.


\subsubsection{Summary of Findings}
Our evaluation demonstrates that the proposed protocol achieves lower aggregation overhead by shifting intensive cryptographic operations to the setup phase,
resulting in reduced latency compared to similar schemes. It also maintains competitive communication cost by eliminating the need for extra communication rounds common in multi-phase decryption protocols, keeping bandwidth usage comparable to or below existing methods. Finally, the use of threshold secret sharing and BS-assisted unmasking provides robust handling of moderate UE and BS dropouts, though overall reliability depends on having a sufficient number of active base stations.

\section{Conclusion and Future Work}

In this paper, we presented a secure aggregation framework for federated learning over 5G networks. By incorporating a single-round protocol with \(t\)-out-of-\(k\) secret sharing and key-homomorphic pseudorandom functions, our approach efficiently safeguards user data while maintaining reliable model updates, even in dynamic and large-scale 5G environments. The empirical results confirm the protocol’s ability to preserve privacy, tolerate adversarial conditions, and adapt to core characteristics of future wireless ecosystems.

Looking ahead, several promising research directions could further enhance this framework. First, an adaptive base station selection mechanism, potentially supported by machine learning, can better handle user mobility by dynamically choosing the most appropriate base stations. 
Second, strategies for share reassignment and revocation may strengthen security, ensuring continuous protection when the network topology changes. 
Third, more advanced or flexible threshold secret sharing schemes-including the integration of zero-knowledge proofs-may bolster resilience against adversarial threats and fluctuating network conditions.
Last, evaluating the protocol on real-world devices remains an important next step, so we leave this as future work and consider it a key step toward practical deployment.

\begin{acks}
This work has been supported by the National Science Foundation (NSF) under Grant No. 2112471, and by the NSF–SNSF joint program under Grant No. ECCS 2444615.
\end{acks}

\bibliographystyle{ACM-Reference-Format}
\balance
\bibliography{references.bib}

\end{document}